\documentstyle[aps,multicol,floats,epsfig,amsfonts,pra]{revtex}

\begin{document}
\draft
\title{Inverted spectroscopy and interferometry for quantum-state
reconstruction of systems with SU(2) symmetry}
\author{C. Brif \cite{pa}\ and\ A. Mann \cite{am}}
\address{Department of Physics, Technion---Israel Institute of
Technology, Haifa 32000, Israel}
\maketitle

\begin{abstract}
We consider how the conventional spectroscopic and interferometric 
schemes can be rearranged to serve for reconstructing quantum states 
of physical systems possessing SU(2) symmetry. The discussed systems 
include a collection of two-level atoms, a two-mode quantized radiation 
field with a fixed total number of photons, and a single laser-cooled 
ion in a two-dimensional harmonic trap with a fixed total number of 
vibrational quanta. In the proposed rearrangement, the 
standard spectroscopic and interferometric experiments are inverted. 
Usually one measures an unknown frequency or phase shift using a 
system prepared in a known quantum state. Our aim is just the inverse 
one, i.e., to use a well-calibrated apparatus with known 
transformation parameters to measure unknown quantum states. 
\end{abstract}


\begin{multicols}{2}

\section{Introduction}

The last few years were marked by an outburst of research devoted to
the problem of reconstruction of quantum states for various physical
systems (see, e.g., Ref.~\cite{BM99} for an extensive list of the
literature on the subject). The problem, as stated already in the 
fifties by Fano \cite{Fano57} and Pauli \cite{Pauli58}, is to 
determine the density matrix $\rho$ from information obtained by a 
set of measurements performed on an ensemble of identically prepared 
systems. Significant theoretical and experimental progress has been 
achieved during the last decade in the reconstruction of quantum 
states of the light field \cite{Leon:book}. Also, numerous works 
were devoted to reconstruction methods for other physical systems.
Most recently, a general theory of quantum-state reconstruction for 
physical systems with Lie-group symmetries was developed \cite{BM99}.

In the present work we consider state-reconstruction methods  
for some quantum systems possessing SU(2) symmetry. The principal
procedure for the reconstruction of spin states was recently 
presented by Agarwal \cite{Agar98}. A similar approach was also 
proposed by Dodonov and Man'ko \cite{DoMa97}, while the basic idea
underlying this method goes back to the pioneering work 
by Royer \cite{Royer}. In brief, one applies a phase-space
displacement [specifically, a rotation in the SU(2) case] to
the initial quantum state and then measures the probability to
find the displaced system in a specific state (the so-called
``quantum ruler'' state). Repeating this procedure with identically 
prepared systems for many phase-space points [many rotation angles
in the SU(2) case], one determines a function on the phase space
(the so-called operational phase-space probability distribution
\cite{Wod84,BKK95,Ban98}). 
In particular, by measuring the population of the ground state, 
one obtains the so-called $Q$ function. The information contained 
in the operational phase-space probability distribution is 
sufficient to completely reconstruct the unknown density matrix of 
the initial quantum state. A general group-theoretical description 
of this method and some examples, including SU(2), are presented 
in Ref.~\cite{BM99}.

The aim of the present paper is to study how the general
state-reconstruction procedure outlined above can be implemented 
in practice for a number of specific physical systems with SU(2) 
symmetry. Three systems are considered: a collection of two-level 
atoms, a two-mode quantized radiation field with a fixed total 
number of photons, and a single laser-cooled ion in a 
two-dimensional harmonic trap with a fixed total number of 
vibrational quanta. We show that a simple rearrangement of 
conventional spectroscopic and interferometric schemes enables one 
to measure unknown quantum states of these systems.

\section{Reconstruction of quantum states for systems with SU(2) 
symmetry}

We start with some basic properties of SU(2) which is the dynamical 
symmetry group for the angular momentum or spin and for many other 
systems (e.g., a collection of two-level atoms, the Stokes 
operators describing the polarization of the quantized light field, 
two light modes with a fixed total photon number, etc.). The su(2)
simple Lie algebra consists of the three operators
$\{J_{x},J_{y},J_{z}\}$, 
\begin{equation}
\label{su2alg}
[J_{p},J_{r}] = i \epsilon_{p r t} J_{t} .
\end{equation}
The Casimir operator is a constant times the unit operator,
${\mathbf{J}}^2 = j(j+1) I$, for any unitary irreducible 
representation of the SU(2) group; so the representations are 
labeled by the single index $j$ that takes the values
$j = 0,1/2,1,3/2,\ldots$. The representation Hilbert space 
${\cal H}_{j}$ is spanned by the complete orthonormal basis 
$\{ |j,\mu\rangle \}$ (where $\mu=j,j-1,\ldots,-j$):
\[
{\mathbf{J}}^2 |j,\mu\rangle = j(j+1) |j,\mu\rangle , \hspace{8mm}
J_z |j,\mu\rangle = \mu |j,\mu\rangle .
\]
In the following we assume that the state $|\psi\rangle$ of the 
system belongs to ${\cal H}_{j}$ (or, for mixed states, that the 
density matrix $\rho$ is an operator on ${\cal H}_{j}$).
Group elements can be 
parametrized using the Euler angles $\alpha,\beta,\gamma$:
\begin{equation}
\label{eq:gEuler}
g=g(\alpha,\beta,\gamma) = e^{ i \alpha J_{z} }  
e^{ i \beta J_{y} } e^{ i \gamma J_{z} } .
\end{equation}

We will employ two very useful concepts: the phase space (which
is the group coset space of maximum symmetry) and the coherent
states (each point of the phase space corresponds to a coherent 
state). For SU(2), the phase space is the unit sphere 
${\Bbb{S}}^2 = {\rm SU}(2) / {\rm U}(1)$, and each coherent
state is characterized by a unit vector \cite{ACGT72,Per}
\begin{equation}
{\bf n} = (\sin\theta \cos\phi, \sin\theta \sin\phi, 
\cos\theta) .
\end{equation}
Specifically, the coherent states $|j;{\bf n}\rangle$ are given by
the action of the group element 
\begin{equation}
\label{eq:oSU2}
g({\bf n}) = e^{- i \phi J_{z} } e^{- i \theta J_{y} }
\end{equation}
on the highest-weight state $|j,j\rangle$:
\begin{eqnarray}
|j;{\bf n}\rangle = g({\bf n}) |j,j\rangle 
& = & \sum_{\mu=-j}^{j} {2j \choose j+\mu}^{1/2} 
\cos^{j+\mu}(\theta/2) \nonumber \\
& & \times \sin^{j-\mu}(\theta/2) e^{- i \mu \phi} |j,\mu\rangle .
\label{jcohstates}
\end{eqnarray}
An important property of the coherent states is the resolution of
the identity:
\begin{equation}
\frac{2j+1}{4\pi} \int_{{\Bbb{S}}^2} d {\bf n}\, |j;{\bf n}\rangle
\langle j;{\bf n}| = I ,
\end{equation}
where $d {\bf n} = \sin\theta\, d \theta\, d \phi$.

A possible procedure for the quantum-state reconstruction is as 
follows \cite{BM99,Agar98,DoMa97}. First, the system, whose initial 
state is described by the density matrix $\rho$, is displaced in 
the phase space:
\begin{equation}
\label{eq:displ}
\rho \rightarrow \rho({\bf n}) = g^{-1}({\bf n}) \rho g({\bf n}) ,
\hspace{8mm} {\bf n} \in {\Bbb{S}}^2 .
\end{equation}
Then one measures the probability to find the displaced system in
one of the states $|j,\mu\rangle$ (e.g., in the highest state
$|j,j\rangle$). This probability 
\begin{equation}
\label{eq:pmu-def}
p_{\mu} ({\bf n}) = \langle j,\mu| \rho({\bf n}) |j,\mu\rangle
\end{equation}
(which is sometimes called the operational phase-space probability 
distribution) can be formally considered as the expectation value 
\begin{equation}
p_{\mu} ({\bf n}) = {\rm Tr}\, [\rho \Gamma_{\mu} ({\bf n})] 
\end{equation}
of the so-called displaced projector 
\begin{equation}
\Gamma_{\mu} ({\bf n}) = g({\bf n}) |j,\mu\rangle 
\langle j,\mu| g^{-1}({\bf n}) .
\end{equation}
Repeating this procedure (with a large number of identically 
prepared systems) for a large number of phase-space points ${\bf n}$,
one can determine the function $p_{\mu} ({\bf n})$.

Knowledge of the function $p_{\mu} ({\bf n})$ is sufficient for the
reconstruction of the initial density matrix $\rho$.
We can use the following expansion for the density matrix (such an
expansion exists for any operator on ${\cal H}_j$):
\begin{equation}
\rho = \sum_{l=0}^{2j} \sum_{m=-l}^{l} {\cal R}_{l m} D_{l m} ,
\hspace{6mm} {\cal R}_{l m} = {\rm Tr}\, (\rho D^{\dagger}_{l m}) .
\end{equation}
Here, $D_{l m}$ are the so-called tensor operators (also known in 
the context of angular momentum as the Fano multipole operators
\cite{Fano53}),
\begin{equation}
D_{l m} = \sqrt{\frac{2l+1}{2j+1}} \sum_{k,q=-j}^{j}
\langle j,k;l,m|j,q \rangle |j,q \rangle \langle j,k| ,
\end{equation}
where $\langle j_{1},m_{1};j_{2},m_{2}|j,m \rangle$ 
are the Clebsch-Gordan coefficients.
Now, one can reconstruct the density matrix by using the relation
\cite{BM99,Agar98}
\begin{equation}
{\cal R}_{l m} = \frac{ \sqrt{(2j+1)/4\pi} }{ 
\langle j,\mu;l,0|j,\mu \rangle } \int_{{\Bbb{S}}^2} d {\bf n}\,
p_{\mu} ({\bf n}) Y_{l m}^{\ast}({\bf n}) ,
\end{equation}
where $Y_{l m}({\bf n})$ are the spherical harmonics.
Other ways to deduce the density matrix from the measured 
probabilities $p_{\mu} ({\bf n})$ were also proposed 
\cite{AmWe99}.

Let us also consider the useful concept of phase-space 
quasiprobability distributions (QPDs). In the SU(2) case, one can
introduce an $s$-parametrized family of the QPDs 
\cite{BM99,Agar81,VaGB89}
\begin{equation}
P({\bf n};s) = \sum_{l=0}^{2j} \sum_{m=-l}^{l} 
\frac{ \sqrt{4\pi/(2j+1)} }{ \langle j,j;l,0|j,j \rangle^{s} }
{\cal R}_{l m} Y_{l m}({\bf n}) .
\end{equation}
For $s=0$, we have the SU(2) equivalent of the Wigner function,
\begin{equation}
W({\bf n}) = \sqrt{ \frac{4\pi}{2j+1} }
\sum_{l=0}^{2j} \sum_{m=-l}^{l} {\cal R}_{l m} Y_{l m}({\bf n}) .
\end{equation}
For $s=1$, we obtain the SU(2) equivalent of the Glauber-Sudarshan
function (also known as Berezin's contravariant symbol), 
$P({\bf n})$, whose defining property is
\begin{equation}
\rho = \frac{2j+1}{4\pi} \int_{{\Bbb{S}}^2} d {\bf n}\, 
P({\bf n}) |j;{\bf n}\rangle \langle j;{\bf n}| .
\end{equation}
The function which is probably the most important for the 
reconstruction problem is the SU(2) equivalent of the Husimi
function (also known as Berezin's covariant symbol),
\begin{equation}
Q({\bf n}) = \langle j;{\bf n}| \rho |j;{\bf n}\rangle ,
\end{equation}
obtained for $s=-1$. As is seen from Eq.~(\ref{eq:pmu-def}),
the function $Q({\bf n})$ gives the probability to find the
displaced system in the highest spin state $|j,j\rangle$,
\begin{equation}
\label{eq:Q-p}
Q({\bf n}) = p_j ({\bf n}) .
\end{equation}
Also, one can see that the probability $p_{-j}(\theta,\phi)$
to find the displaced system in the lowest spin state 
$|j,-j\rangle$ is equal to $Q(\theta+\pi,\phi)$.
More generally, any one of the QPDs can be reconstructed using
the relation \cite{BM99}
\begin{eqnarray}
&& P({\bf n};s) = \frac{2j+1}{4\pi} \int_{{\Bbb{S}}^2} d {\bf n}'\,
K_{\mu,s}^{-}({\bf n}, {\bf n}')\, p_{\mu} ({\bf n}') , \\
&& K_{\mu,s}^{-}({\bf n}, {\bf n}') = \sum_{l=0}^{2j} 
\frac{2l+1}{2j+1} \frac{ \langle j,j;l,0|j,j \rangle^{-s} }{ 
\langle j,\mu;l,0|j,\mu \rangle } P_l ( {\bf n} \cdot {\bf n}' ) ,
\end{eqnarray}
where $P_l (x)$ are the Legendre polynomials. 
For $s=-1$ and $\mu=j$ we recover the relation (\ref{eq:Q-p}).

\section{General description of experimental schemes}

\subsection{Spectroscopy and interferometry}

Quantum transformations which constitute the basic operations
in spectroscopic and interferometric measurements can be
conveniently described as rotations in an abstract 3-dimensional
space. In this description, the system is characterized by the
vector ${\mathbf{J}} = (J_x , J_y , J_z)^T$, where the three
operators $J_x$, $J_y$, and $J_z$ satisfy the su(2) algebra
(\ref{su2alg}).

A spectroscopic or interferometric process is usually described
in the Heisenberg picture as a unitary transformation
\begin{equation}
{\mathbf{J}}_{\mathrm{out}} = 
U(\vartheta_1,\vartheta_2,\varphi) {\mathbf{J}} 
U^{\dagger}(\vartheta_1,\vartheta_2,\varphi)
= {\mathsf{U}}(\vartheta_1,\vartheta_2,\varphi) {\mathbf{J}} ,
\end{equation}
where ${\mathsf{U}}(\vartheta_1,\vartheta_2,\varphi)$ is a $3 \times 3$ 
transformation (rotation) matrix, and $\vartheta_1$, $\vartheta_2$, 
$\varphi$ are transformation parameters (rotation angles). 
A standard transformation consists of three steps: 
\begin{enumerate}
\item[(i)] rotation around the $\hat{\mathbf{y}}$ axis by $\vartheta_1$, 
with the transformation matrix ${\mathsf{R}}_y (\vartheta_1)$,
\item[(ii)] rotation around the $\hat{\mathbf{z}}$ axis by $\varphi$, 
with the transformation matrix ${\mathsf{R}}_z (\varphi)$,
\item[(iii)] rotation around the $\hat{\mathbf{y}}$ axis by 
$\vartheta_2$, with the transformation matrix 
${\mathsf{R}}_y (\vartheta_2)$.
\end{enumerate}
The overall transformation performed on ${\mathbf{J}}$ is
\begin{equation}
{\mathsf{U}}(\theta,\phi) = {\mathsf{R}}_y (\vartheta_2)
{\mathsf{R}}_z (\varphi) {\mathsf{R}}_y (\vartheta_1) .
\end{equation}
This transformation is slightly more general than those routinely
made in spectroscopy and interferometry. The usual choice is
$\vartheta_2 = -\vartheta_1 = \pm \pi/2$, so 
${\mathsf{U}} = {\mathsf{R}}_x (\pm \varphi)$, respectively,
while $\varphi$ is the parameter to be estimated in the experiment.
In the Schr\"{o}dinger picture, the density matrix of the system
transforms as
\begin{equation}
  \label{eq:Srot}
\rho_{\mathrm{out}} = U^{\dagger}(\vartheta_1,\vartheta_2,\varphi) 
\rho U(\vartheta_1,\vartheta_2,\varphi) , 
\end{equation}
where the transformation operator is
\begin{equation}
  \label{eq:Uoperator}
U(\vartheta_1,\vartheta_2,\varphi) = e^{ i \vartheta_1 J_y }
e^{ i \varphi J_z } e^{ i \vartheta_2 J_y } .
\end{equation}

Now, the aim is to measure the value of $\varphi$ which is 
proportional to the transition frequency in a spectroscopic 
experiment or to the optical path difference between the two arms of 
an interferometer.
The information on $\varphi$ is inferred from the measurement of
the observable $J_z$ at the output. The quantum uncertainty
in the estimation of $\varphi$ is
\begin{equation}
\label{eq:uncert}
\Delta \varphi = \frac{\Delta J_{z \mathrm{out}} }{| \partial
\langle J_{z \mathrm{out}} \rangle/ \partial \varphi |} ,
\end{equation}
where the expectation values are taken over the initial quantum
state of the system. This state is assumed to be known, so one
can estimate the value of $\varphi$ and the corresponding 
uncertainty. 

\subsection{Reconstruction of the initial state}

In this paper we consider how to use the spectroscopic or 
interferometric arrangement for the inverse purpose, i.e., for the
measurement of an unknown initial quantum state by means of a large 
number of transformations with known parameters.

As discussed in Sec.~II, the first part of the reconstruction 
procedure is the phase-space displacement of Eq.~(\ref{eq:displ}).
With the phase space being the sphere, this displacement is just
a rotation produced by the operator $g({\bf n})$ of 
Eq.~(\ref{eq:oSU2}). Now, compare this rotation with the one
made during a spectroscopic or interferometric experiment, as
given by Eqs.~(\ref{eq:Srot}) and (\ref{eq:Uoperator}). 
One can immediately conclude that the phase-space displacement 
needed for the SU(2) state-reconstruction procedure can be neatly 
implemented by means of the spectroscopic and interferometric 
techniques. One only needs to omit the first rotation (i.e., take 
$\vartheta_1 = 0$), and recognize the two spherical angles as:
\begin{equation}
\theta = -\vartheta_2 , \hspace{8mm}
\phi = -\varphi .
\end{equation}
After the rotation $g({\bf n})$ is made, one should
measure the probability $p_{\mu}({\bf n})$ to find the displaced
system in the state $|j,\mu\rangle$. Perhaps the most convenient 
choice is to measure the population of the lowest state 
$|j,-j\rangle$, which is usually the ground state of the system
(e.g., this state corresponds to the case where all the atoms are 
unexcited; in the atomic case such a measurement can be made by 
monitoring the resonant fluorescence for an auxiliary dipole 
transition). 
This procedure should be repeated for many phase-space points
${\bf n}$ with a large number of identically prepared systems, 
thereby determining the function $p_{\mu}({\bf n})$ (e.g., for
$\mu = j$ or $\mu = -j$). According to the formalism presented
in Sec.~II, this information is sufficient to reconstruct the 
initial quantum state.  

\section{Collections of two-level atoms}

In Ramsey spectroscopy \cite{Ramsey} one deals with a collection 
of $N$ two-level systems (usually atoms or ions) interacting with 
classical light fields. One can equivalently describe this physical 
situation as the interaction of $N$ spin-$\frac{1}{2}$ particles 
with classical magnetic fields. Denoting by ${\mathbf{S}}_i$ the 
spin of $i$th particle, one can use the collective spin operators:
\begin{equation}
{\mathbf{J}} = \sum_{i=1}^{N} {\mathbf{S}}_i .
\end{equation}
The orthonormal basis $\{ |j,\mu\rangle \}$ consists of the symmetric 
Dicke states \cite{Dicke54}:
\begin{equation}
|j,\mu\rangle = {N \choose p}^{-1/2} \sum \prod_{k=1}^{p} 
|+\rangle_{l_k} \prod_{l \neq l_k} |-\rangle_{l} ,
\end{equation}
where $|+\rangle_l$ and $|-\rangle_l$ are the upper and lower states,
respectively, of the $l$th atom, and the summation is over all
possible permutations of $N$ atoms. If only symmetric states
are considered, then the ``cooperative number'' $j$ is equal to
$N/2$ and $p = \mu+j$ is just the number of excited atoms.
Usually the atoms (ions) are far enough apart so their wave functions
do not overlap and the direct dipole-dipole coupling or other direct
interactions between the atoms may be neglected.

In the spin formulation (see, e.g., Ref.~\cite{Wine94} for a very good 
description), the magnetic moment $\bbox{\mu} = \mu_0 {\mathbf{S}}$ 
is associated with each particle. If a uniform external magnetic 
field ${\mathbf{B}}_0 = B_0 \hat{\mathbf{z}}$ 
is applied, the Hamiltonian for each particle is given by
\begin{equation}
H_0 = - \bbox{\mu} \cdot {\mathbf{B}}_0 = \hbar \omega_0 S_z ,
\end{equation}
where $\hbar \omega_0 = -\mu_0 B_0$ is the separation in energy
between the two levels. The corresponding Heisenberg equation for
the collective spin operator is
\begin{equation}
\partial {\mathbf{J}}/ \partial t = \bbox{\omega}_0 \times 
{\mathbf{J}} ,
\end{equation}
where $\bbox{\omega}_0 = \omega_0 \hat{\mathbf{z}}$. Then one
applies the so-called clock radiation which is a classical field
of the form
\begin{equation}
{\mathbf{B}}_{\perp} = B_{\perp} \left( \hat{\mathbf{y}} \cos \omega t
- \hat{\mathbf{x}} \sin \omega t \right) ,
\end{equation}
where $\omega \simeq \omega_0$ and we assume $\omega_0 > 0$.
In the reference frame that rotates at frequency $\omega$, the
collective spin ${\mathbf{J}}$ interacts with the effective field
\begin{equation}
{\mathbf{B}} = B_r \hat{\mathbf{z}} + B_{\perp} \hat{\mathbf{y}} ,
\end{equation}
where $B_r = B_0 (\omega_0 -\omega)/\omega_0$. In the rotating 
frame, the Hamiltonian is 
$H = - \mu_0 {\mathbf{J}} \cdot {\mathbf{B}}$, and the Heisenberg 
equation for $\mathbf{J}$ is 
\begin{equation}
\label{eq:Jrot}
\partial {\mathbf{J}}/ \partial t = \bbox{\omega}' \times 
{\mathbf{J}} ,
\end{equation}
where $\bbox{\omega}' = (\omega_0 - \omega) \hat{\mathbf{z}}
+ \omega_{\perp} \hat{\mathbf{y}}$ and 
$\omega_{\perp} = -\mu_0 B_{\perp}/\hbar$.

The Ramsey method breaks the evolution time of the system into three
parts. In the first part $B_{\perp}$ is nonzero and constant with 
value $B_{1}$ during the time interval $0 \leq t \leq t_{\vartheta}$. 
During this period (the first Ramsey pulse), 
${\mathbf{B}} = B_r \hat{\mathbf{z}} + B_{1} \hat{\mathbf{y}}
\simeq B_{1} \hat{\mathbf{y}}$, where we made the assumption
$|B_{1}| \gg |B_{r}|$, i.e., $|\omega_{1}| \gg |\omega_0 - \omega|$,
with $\omega_{1} = -\mu_0 B_{1}/\hbar$. Therefore, in the rotating 
frame of Eq.~(\ref{eq:Jrot}), $\mathbf{J}$ rotates around the 
$\hat{\mathbf{y}}$ axis by the angle 
$\vartheta_1 = \omega_{1} t_{\vartheta}$.
During the second period, of duration $T$, 
(usually $T \gg t_{\vartheta}$), $B_{\perp}$ is zero, so 
${\mathbf{B}} = B_r \hat{\mathbf{z}}$, and $\mathbf{J}$ rotates around 
the $\hat{\mathbf{z}}$ axis by the angle 
$\varphi = (\omega_0 - \omega) T$. The third period is exactly as the
first one but with the field $B_{\perp} = B_{2}$ and the corresponding
angular frequency $\omega_{2} = -\mu_0 B_{2}/\hbar$.
This gives a rotation around the $\hat{\mathbf{y}}$ axis by the
angle $\vartheta_2 = \omega_{2} t_{\vartheta}$. These three Ramsey 
pulses provide the rotations we described in Sec.~III~A (usually,
$\vartheta_1 = \vartheta_2 = \pi/2$).

The aim of spectroscopic experiments is to measure the transition 
frequency $\omega_0$ (which is equivalent to the measurement of
$\varphi$, as $\omega$ and $T$ are determined by the experimenter).  
Usually, one measures the number of atoms in the upper state
$|+\rangle$,
\begin{equation}
N_{+ {\rm out}} = J_{z {\rm out}} + N/2 ,
\end{equation}
and thus obtains the information about the angle $\varphi$ or,
equivalently, about the frequency $\omega_0$. Of course, in order to 
infer this information one should know the initial quantum state of 
the system. The measurement sensitivity, as seen from 
Eq.~(\ref{eq:uncert}), also depends on the initial quantum state.

In the state-reconstruction procedure, the first Ramsey pulse should
be omitted, while the second and third pulses produce the desired
phase-space displacement $g^{\dagger}({\bf n})$. After the displacement 
is completed, one should measure the probability to find the system in 
one of the states $|j,\mu\rangle$, for example, measure the population 
of the ground state $|j,-j\rangle$ or of the most excited state 
$|j,j\rangle$.
This measurement can be made by driving a dipole transition to an
auxiliary atomic level and then observing the resonance fluorescence.
The phase space is scanned by repeating the measurement with many
identically prepared systems for various durations of the
Ramsey pulses, $T$ and $t_{\vartheta}$. Of course, the apparatus
should be first calibrated by measuring the transition frequency
$\omega_0$.

\section{Two-mode light fields}

The basic device employed in a passive optical interferometer is
a beam splitter (a partially transparent mirror). A Mach-Zehnder
interferometer consists of two beam splitters and its operation
is as follows. Two light modes (with boson annihilation operators
$a_1$ and $a_2$) are mixed by the first beam splitter,
accumulate phase shifts $\varphi_1$ and $\varphi_2$, respectively,
and then they are once again mixed by the second beam splitter.
Photons in the output modes are counted by two photodetectors.
In fact, a Michelson interferometer works in the same way, but
due to its geometric layout the two beam splitters may coincide.

Each beam splitter has two input and two output ports.
Let ${\bf a} = (a_1 , a_2)^T$ and ${\bf b} = (b_1 , b_2)^T$ be the 
column-vectors of the boson operators of the input and output modes,
respectively. Then, in the Heisenberg picture, the action of
the beam splitter is described by the transformation
\begin{equation}
{\bf b} = {\sf B} {\bf a} ,
\end{equation}
where ${\sf B}$ is a $2 \times 2$ matrix. For a lossless beam
splitter ${\sf B}$ must be unitary, thereby assuring the energy
(photon number) conservation. A possible form of ${\sf B}$ is
\begin{equation}
\label{eq:BSmatrix}
{\sf B}(\vartheta) = \left( \begin{array}{cc} 
\cos (\vartheta/2) & -\sin (\vartheta/2) \\
\sin (\vartheta/2) & \cos (\vartheta/2) \end{array} \right) ,
\end{equation}
with $T = \cos^2 (\vartheta/2)$ and $R = \sin^2 (\vartheta/2)$
being the transmittance and reflectivity, respectively.
When the two light modes accumulate phase shifts $\varphi_1$ and 
$\varphi_2$, respectively, the corresponding transformation is
\begin{equation}
\label{eq:PSmatrix}
{\bf b} = {\sf P} {\bf a} , \hspace{8mm}
{\sf P} = \left( \begin{array}{cc} 
e^{ i \varphi_1 } & 0 \\ 0 & e^{ i \varphi_2 } \end{array} \right) .
\end{equation}

The group-theoretic description of the interferometric process 
\cite{YMK86} is based on the Schwinger realization of the su(2) 
algebra:
\begin{eqnarray}
  \label{eq:Schwinger}
& & J_x = 
( a_1^{\dagger} a_2 + a_2^{\dagger} a_1 )/2 , \nonumber \\
& & J_y = 
- i ( a_1^{\dagger} a_2 - a_2^{\dagger} a_1 )/2 , \\
& & J_z = 
( a_1^{\dagger} a_1 - a_2^{\dagger} a_2 )/2 . \nonumber
\end{eqnarray}
Actions of the interferometer elements (mixing by the beam splitters 
and phase shifts) can be represented as rotations of the column-vector 
${\bf J} = ( J_x , J_y , J_z )^T$.
The beam-splitter transformation of Eq.~(\ref{eq:BSmatrix}) is 
represented by rotation ${\sf R}_y (\vartheta)$ around the 
$\hat{\mathbf{y}}$ axis by the angle $\vartheta$, and the phase shift 
of Eq.~(\ref{eq:PSmatrix}) is represented by rotation 
${\sf R}_z (\varphi)$ around the $\hat{\mathbf{z}}$ axis by the angle 
$\varphi = \varphi_2 - \varphi_1$.
Now, if the transmittances of the two beam splitters are 
$T_1 = \cos^2 (\vartheta_1 /2)$ and $T_2 = \cos^2 (\vartheta_2 /2)$,
respectively, then the interferometer action is given by the
three rotations described in Sec.~III~A. (Usually, one uses 50-50
beam splitters, so $\vartheta_1 = -\vartheta_2 = \pi/2$.)

Interferometers are constructed to measure the relative phase
shift $\varphi$, which is proportional to the optical path difference
between the two arms. Usually, one measures the difference between 
the photocurrents due to the two output light beams. This quantity is
proportional to the photon-number difference at the output,
$q_{\mathrm{out}} = 2 J_{z \mathrm{out}}$. If the input state of
light is known, then the measurement of $q_{\mathrm{out}}$ can
be used to infer the phase shift $\varphi$ and estimate the
measurement error due to the quantum fluctuations of the light
field.

A simple calculation gives ${\bf J}^2 = (N/2)(1+N/2)$, where
$N = a_1^{\dagger} a_1 + a_2^{\dagger} a_2$ is the total number
of photons in the two modes. If $N$ has a fixed value for the input
state of the two-mode light field, then this state belongs to the 
Hilbert space ${\cal H}_j$ of a specific SU(2) representation with 
$j = N/2$. Because $N$ is the SU(2) invariant, this state will remain 
in ${\cal H}_j$ during the interferometric process. 
Such input states of the two-mode light field can be reconstructed 
using a rearrangement of the interferometric scheme, according to the 
general procedure described in Secs.~II and III~B.

The phase-space displacement $g^{\dagger}({\bf n})$ needed for the 
state-reconstruction procedure can be implemented by using an 
interferometer without the first beam splitter. Then one should 
measure the probability $p_{\mu}({\bf n})$ to find the output light 
in one of the states $|j,\mu\rangle$. Note that these states are 
given by
\begin{equation}
\label{eq:staterel}
|j,\mu\rangle = |j+\mu \rangle_1 \otimes |j-\mu \rangle_2
\end{equation}
in the terms of the Fock states of the two light modes.
So, $\mu$ is just one half of the photon-number difference 
measured at the output. Averaging over many measurements, one
obtains the probabilities $p_{\mu}({\bf n})$.
For example, $p_{j}({\bf n})$ is the probability that all photons 
exit in the first output beam while the number of photons in the 
second output beam is zero.
The measurement should be repeated with identically prepared input
light beams for many phase-space displacements. This means that one 
needs a well-calibrated apparatus which can be tuned for various 
values of the relative phase shift $\varphi$. These phase shifts 
can be conveniently produced by moving a mirror with a precise 
electro-mechanical system. Various values of the angle 
$\vartheta_2$ can be realized using a collection of partially
transparent mirrors with different reflectivities for the second
beam splitter. An alternative possibility is to use the dependence
of the reflectivity on the angle of incidence for light polarized 
in the plane of incidence.

In general, the state reconstruction for two-mode light fields is
a tedious task, because the corresponding Hilbert space is very 
large \cite{KWV95,RMAL96,OWV97,Richter97,PTKJ97}.
Obviously, this task can be greatly simplified for the subclass of 
two-mode states with a fixed total number of photons, by means of
the reconstruction method presented here. However, this method is
in principle suitable also for other two-mode states as well.
In general, the whole Hilbert space of the two-mode system can
be decomposed as
\begin{equation}
\label{eq:decomp}
{\cal H} = \bigoplus_{j} {\cal H}_j .
\end{equation}
The method of inverted interferometry enables one to reconstruct 
the part of the density matrix corresponding to each irreducible 
subspace ${\cal H}_j$.
One case for which our method is applicable is the subclass of 
states, whose density matrices are block-diagonal in terms of the 
decomposition (\ref{eq:decomp}).
This means that the corresponding operator can be written as
\begin{equation}
\rho = \sum_{j} \rho_j ,
\end{equation}
where $\rho_j$ is an operator on ${\cal H}_j$. Each component 
$\rho_j$ evolves independently during the phase-space displacement; 
hence the state of the whole system can be measured by  
reconstructing all invariant components $\rho_j$.
The other case for which our method works is the subclass of pure 
states,
\begin{equation}
|\psi\rangle = \sum_{j} |\psi_j\rangle , \hspace{8mm}
|\psi_j\rangle = \sum_{\mu = -j}^{j} c_{j\mu} |j,\mu\rangle .
\end{equation}
Then the density matrix can be written as
\begin{equation}
\label{eq:pure-decomp}
\rho = \sum_{j} | \psi_j \rangle \langle \psi_j | + 
\sum_{j \neq j'} |\psi_j \rangle \langle \psi_{j'} | .
\end{equation}
The populations of the states $|j,\mu\rangle$ are unaffected by
the second term in (\ref{eq:pure-decomp}), and one can reconstruct
all invariant components 
$\rho_j = | \psi_j \rangle \langle \psi_j |$.
This gives information about the state $|\psi\rangle$ of the whole 
system, except for relative phases between different 
$| \psi_j \rangle$. From the technical point of view,
each measurement of the photon-number difference $2 \mu$, needed 
to determine the probabilities $p_{\mu}({\bf n})$, should be
accompanied by a measurement of the photon-number sum $N = 2j$, 
in order to determine to which invariant subspace ${\cal H}_j$ 
does the detected value of $\mu$ correspond. 
Consequently, one needs to make many more measurements, in order 
to accumulate enough data for each value of $j$. 
A technical problem is that quantum efficiencies of realistic 
photodetectors are always less then unity.
While this problem is not too serious for the measurement of the
photon-number difference (as long as both detectors have the
same efficiency), it puts a serious limitation on the accuracy
of the measurement of the total number of photons.

\section{Two-dimensional vibrations of a trapped ion}

As was recently demonstrated by Wineland \emph{et al.} \cite{Wine98},
a single laser-cooled ion in a harmonic trap can be used to simulate 
various interactions governing many well-known optical processes.
In particular, one can simulate transformations produced by elements 
of a Mach-Zehnder optical interferometer.

Consider a single ion confined in a two-dimensional harmonic trap, 
with angular frequencies of oscillations in two orthogonal
directions $\Omega_1$ and $\Omega_2$. Two internal states
of the ion, $|+\rangle$ and $|-\rangle$, are separated in energy
by $\hbar \omega_0$. The internal and motional degrees of freedom
can be coupled by applying classical laser beams, with 
electric fields of the form
\[ {\bf E}({\bf x},t) = {\bf E}_0 \cos ({\bf k} \cdot {\bf x}
- \omega t + \Phi) . \]
For example, one can apply two laser beams to produce stimulated
Raman transitions. We denote by $\omega = \omega_1 - \omega_2$,
${\bf k} = {\bf k}_1 - {\bf k}_2$, and $\Phi = \Phi_1 -  \Phi_2$
the differences between the angular frequencies, the wave 
vectors, and the phases, respectively, of the two applied fields.
Then, in the rotating-wave approximation, the interaction 
Hamiltonian reads
\begin{equation}
H_I = \hbar \kappa \exp[ i ({\bf k} \cdot {\bf x} - \delta t + 
\Phi) ] + {\rm H.c.} ,
\end{equation}
where $\delta = \omega-\omega_0$ is the frequency detuning,
${\bf x}$ is the ion's position relative to its equilibrium, and
$\kappa$ is the coupling constant (the Rabi frequency). Each of 
the two modes of the ion's motion can be modelled by a quantum 
harmonic oscillator:
\begin{equation}
x_r = x_{0 r} (a_r + a_r^{\dagger}), \hspace{6mm}
x_{0 r} = \sqrt{ \hbar/(2 M \Omega_r) } ,
\end{equation}
where $r=1,2$ and $M$ is the ion's mass. Also, let 
$\eta_r = k_r x_{0 r}$ ($r=1,2$) be the Lamb-Dicke parameters for the 
two oscillatory modes. 
It is convenient to use the interaction picture for the ion's motion: 
\begin{eqnarray}
\tilde{H}_I & = & \exp( i H_0 t/\hbar) H_I \exp(- i H_0 t/\hbar) 
\nonumber \\
& = & \hbar \kappa e^{ i (\Phi - \delta t)} \prod_{r=1,2}
\exp[ i \eta_r (\tilde{a}_r + \tilde{a}_r^{\dagger}) ] + {\rm H.c.}, 
\label{eq:Ham2}
\end{eqnarray}
where $H_0$ is the free Hamiltonian for the ion's motion,
\begin{equation}
H_0 = \hbar \Omega_1 \left( a_1^{\dagger} a_1 + \mbox{$\frac{1}{2}$}
\right) + \hbar \Omega_2 \left( a_2^{\dagger} a_2 + 
\mbox{$\frac{1}{2}$} \right) ,
\end{equation}
and $\tilde{a}_r = a_r \exp(- i \Omega_r t)$, $r=1,2$.

If the coupling constant $\kappa$ is small enough and $\Omega_1$ and 
$\Omega_2$ are incommensurate, one can resonantly excite only one 
spectral component of the possible transitions. For a particular 
resonance condition $\delta = \Omega_2 - \Omega_1$ (and in the 
Lamb-Dicke limit of small $\eta_1$ and $\eta_2$), the product in 
Eq.~(\ref{eq:Ham2}) will be dominated by the single term
$( i \eta_1 a_1 )( i \eta_2 a_2^{\dagger} )$. Therefore, one
obtains
\begin{equation}
\label{eq:H-bs}
\tilde{H}_I \approx -\hbar \kappa \eta_1 \eta_2 \left( e^{ i \Phi} 
a_1 a_2^{\dagger} + e^{- i \Phi} a_1^{\dagger} a_2 \right).
\end{equation}
Returning to the Schr\"{o}dinger picture, the total evolution
operator reads:
\begin{eqnarray}
U(t) & = & \exp(- i H_0 t/\hbar) \exp(- i \tilde{H}_I t/\hbar)
\nonumber \\
& = & \exp[- i (\Omega_1 + \Omega_2)(N+1) t/2 ]
\exp[ i (\Omega_2 - \Omega_1) J_z t ] \nonumber \\
& & \times \exp( 2  i \kappa \eta_1 \eta_2 J_{\Phi} t ) .
\label{eq:evolution}
\end{eqnarray}
Here, $N = a_1^{\dagger} a_1 + a_2^{\dagger} a_2$ is the total number
of vibrational quanta in the two modes, 
$J_{\Phi} = J_x \cos\Phi + J_y \sin\Phi$, and we used the Schwinger 
realization (\ref{eq:Schwinger}) for the SU(2) generators. 

Now, let us consider only such motional states of the ion for which
$N$ has a fixed value, i.e., which belong to the irreducible Hilbert 
space ${\cal H}_j$ (with $j = N/2$). For these states,
the first exponent in (\ref{eq:evolution}) will just produce an
unimportant phase factor and can be omitted. Clearly, the evolution 
operator (\ref{eq:evolution}) can be used to simulate the action of 
an optical interferometer, with two vibrational modes of a trapped 
ion employed instead of two light beams. In order to simulate the
action of a beam splitter, one should apply the interaction
(\ref{eq:H-bs}) during time $t_{\theta}$ and ensure that
$|2 \kappa \eta_1 \eta_2| \gg |\Omega_2 - \Omega_1|$, so the effect
of the free evolution can be neglected. Then, for $\Phi = \pi/2$,
the evolution operator reads
\begin{equation}
\label{eq:Utheta}
U_y (\theta) = \exp( i \theta J_y ) , \hspace{8mm}
\theta = 2 \kappa \eta_1 \eta_2 t_{\theta} .
\end{equation}
A relative phase shift between the two modes can be produced
just by using the free evolution, i.e., with no external laser 
fields applied. Letting the system evolve freely during time $T$,
one obtains
\begin{equation}
\label{eq:Uphi}
U_z (\phi) = \exp( i \phi J_z ) , \hspace{8mm}
\phi = (\Omega_2 - \Omega_1) T .
\end{equation}

It is obvious that applying consequently the transformations
(\ref{eq:Uphi}) and (\ref{eq:Utheta}) one will produce the 
phase-space displacement $g^{\dagger}({\bf n})$, employed in the 
state-reconstruction procedure. The whole phase space can be 
scanned by repeating the procedure with identically prepared
systems for various durations $T$ and $t_{\theta}$. Each 
phase-space displacement should be followed by the measurement of 
the probability $p_{\mu}({\bf n})$ to find the system in one of
the states $|j,\mu\rangle$. For example, $p_{j}({\bf n})$ is
the probability that the first oscillatory mode is excited to the
$N$th level ($N = 2j$) while the second mode is in the ground 
state. Such a measurement can be made with the method used
recently by the NIST group \cite{Leibfr} to reconstruct the 
one-dimensional motional state of a trapped ion. 
The principle of this method is as follows. One of the 
oscillatory modes is coupled to the internal transition 
$|+\rangle \leftrightarrow |-\rangle$. This is done by applying
one classical laser field, so single-photon transitions are
excited. This results in an interaction of the Jaynes-Cummings
type \cite{JC63} between the oscillatory mode and the internal 
transition. Then the population $P_{-}(t)$ of the lower internal 
state $|-\rangle$ is measured for various values of the
interaction time $t$ (as we already mentioned, this measurement 
can be made by monitoring the resonant fluorescence produced in 
an auxiliary dipole transition). If $|-\rangle$ is the internal 
state at $t=0$, then the signal averaged over many measurements 
is 
\[
P_{-}(t) = \frac{1}{2} \left[ 1 + 
\sum_{n=0}^{\infty} P_{n} \cos(2 \Omega_{n,n+1} t)
e^{-\gamma_{n} t} \right] ,
\]
where $\Omega_{n,n+1}$ are the Rabi frequencies and $\gamma_{n}$ 
are the experimentally determined decay constants. This
relation allows one to determine the populations $P_{n}$ 
of the motional eigenstates $|n\rangle$. By virtue of 
Eq.~(\ref{eq:staterel}), this gives the populations $p_{\mu}$ of 
the SU(2) states $|j,\mu\rangle$ (with $\mu = n-j$ for the first
mode and $\mu = j-n$ for the second mode). For example, $p_{-j}$
and $p_{j}$ are given by $P_0$ for the first and second modes,
respectively.

\section{Conclusions}

In this paper we presented practical methods for the reconstruction 
of quantum states for a number of physical systems with SU(2) 
symmetry. All these methods employ the same basic idea---the 
measurement of displaced projectors---which in principle is 
applicable to any system possessing a Lie-group symmetry. Practical 
realizations, of course, vary for different physical systems. 
In our approach, we exploited the fact that transformations applied
in conventional spectroscopic and interferometric schemes are, from
the mathematical point of view, just rotations. In the context of
the SU(2) group, these rotations constitute phase-space 
displacements needed to implement a part of the reconstruction
procedure. Therefore, the spectroscopic and interferometric
measurements can be easily rearranged in order to enable one to 
determine unknown quantum states for an ensemble of identically 
prepared systems. As the spectroscopic and interferometric
measurements are known for their high accuracy, we hope that the
corresponding rearrangements will allow accurate reconstructions 
of unknown quantum states.

\acknowledgements

This work was supported by the Fund for Promotion of Research 
at the Technion and by the Technion VPR Fund.

\end{multicols}

\end{document}